\documentclass{article}
 \makeatletter
    
    \@addtoreset{equation}{section}
  \makeatother
\usepackage{amsmath}
\usepackage[psamsfonts]{amssymb}
\usepackage{epsf}
\usepackage{latexsym}
\usepackage{graphicx}

\begin{document}
\begin{center}
{ \large \bf Causal Dynamical Triangulation \\
with \\ 
 Extended Interactions in 1$+$1 Dimensions }

 \vspace{30pt}

{\sl Hiroyuki\ Fuji}$\,^{a}$,
{\sl Yuki\  Sato}$\,^{a,b}$ and
{\sl Yoshiyuki\ Watabiki}$\,^{c}$

\vspace{24pt}

{\footnotesize

$^a$~ Nagoya University,\\
Dept. of Physics, Graduate School of Science, \\
Furo-cho, Chikusa-ku, Nagoya 464-8602, Japan\\
{email: fuji@th.phys.nagoya-u.ac.jp}\\

\vspace{10pt}

$^b$~The Niels Bohr Institute, Copenhagen University\\
Blegdamsvej 17, DK-2100 Copenhagen \O , Denmark.\\
{ email: ysato@nbi.dk }\\

\vspace{10pt}

$^c$~Tokyo Institute of Technology,\\ 
Dept. of Physics, High Energy Theory Group,\\ 
2-12-1 Oh-okayama, Meguro-ku, Tokyo 152-8551, Japan\\
{email: watabiki@th.phys.titech.ac.jp}\\

}
\vspace{24pt}

\end{center}

\begin{center}
{\bf Abstract}
\end{center}

We study the Causal Dynamical Triangulation (CDT) with extended interactions in 1$+$1 dimensions
applying the method in the non-critical string field theory (SFT) constructed by Ishibashi and Kawai.
For this model, we solve Schwinger-Dyson's equation (SDE) for disk amplitude perturbatively, and 
find a matrix model in the continuum limit reproducing the SDE in the non-critical SFT approach
as the loop equation.  

\pagebreak
\section{Introduction}
As a consequence of general relativity, ``un-countable'' lots of physics about Universe have been uncovered. Now, we have to go beyond the theory and into the quantum realm, i.e. \textit{quantum gravity}.   
However, it has been known that there is a difficulty in the case that we extend general relativity to quantum gravity. Namely, general relativity is not renormalizable at least perturbatively. 

As a candidate to overcome such a serious problem, a kind of non-perturbative method has been proposed, which is called \textit{Euclidean Dynamical Triangulation} (EDT). In EDT, discretizing Euclidean space-time by simplices having the lattice spacing $a$ as each side length, we can carry out the Euclidean gravitational path-integral non-perturbatively. An important point here is that the lattice spacing $a$ is about the inverse energy cut-off $\Lambda_{\text{grav}}$. Unfortunately, in EDT, no reasonable classical space-time has been found in 4 dimensions, and what has been found are only the skinny polymer-like geometry or the dense crumpled geometry, which has been calculated with the help of Monte-Carlo simulations. This is because the geometries based on EDT is too ``wild'' to handle. Alternatively speaking, infinite numbers of baby universes are produced in this approach. However, EDT had a great deal of success in 2 dimensions. In the suitable continuum limit, physical quantities such as several critical exponents and correlation functions in EDT realize those of the so-called \textit{quantum Liouville theory}. Furthermore, a dual expression of EDT has been found, and it is called the \textit{matrix model}. Utilizing the powerfulness of the matrix model, conformal matters realized in the so-called \textit{minimal model} have been successfully included in EDT approach.

In this line of thought, a kind of breakthrough has been casted out, which is known as \textit{Causal Dynamical Triangulation} (CDT) \cite{a13}. In CDT approach, the path-integral of dynamically triangulated geometries can be done non-perturbatively under the two new additional restrictions. First, one gives the Lorentzian signature to simplices. Second, the time-foliation structure is imposed. In this approach, our de-Sitter universe can be ``realized'' in $3+1$ dimensions \cite{a0}. Furthermore, in $1+1$ dimensions physical quantities, say disk amplitude, can be solved analytically \cite{a13}. An outstanding feature of the pure CDT is that no baby universe is allowed according to the non-anomalous scaling dimension of time.
Related to the fact above, for instance the Hausdorff dimension $d_{H}$ in the (1+1)-dimensional setup is not anomalous, $d_{H}=2$, compared to that in EDT, $d_{H}=4$. 
This is one of attractive traits of CDT. 

The CDT approach really restricts the configurations of geometries to the causal ones a priori, but in fact we do not understand whether or not we should exclude the contributions from the baby universes and furthermore from the geometries with different space-time topologies. 
Focusing on the (1+1)-dimensional case, CDT has been extended to the one including topology changing processes within the criterion that the scaling behavior does not change, i.e. the causal geometries are still dominant, via the non-critical String Field Theory (SFT) \cite{a3}. In addition, the matrix model expression for the non-critical SFT based on CDT has been found \cite{a9}. Such extended models including baby universes and topology changes are called \textit{Generalized CDT's} (GCDT). As for the matter-coupled CDT's, there is not any analytical tool to calculate even in 1+1 dimensions. 
From the lessons based on the subsequent works in CDT, it can be said that the dominance of causal geometries, characterized by the fact that space and time have the same scaling dimension, prevents the stampede of geometries. Alternatively speaking, the causality makes geometries obedient to handle. If the CDT approach is on a correct direction as quantum gravity, this may be a pretty nice property. 

In this paper, to read off some hidden traits of CDT, we quest for possibilities to extend the GCDT approach without changing the scaling dimensions of space and time in 1+1 dimensions. We actually extend GCDT applying the method in the non-critical SFT techniques in \cite{a1} and \cite{a2}. We solve the Schwinger-Dyson's equation (SDE) for disk amplitude in our model by the perturbation w.r.t. the string coupling constant. Moreover, we define the corresponding matrix model in the continuum limit. In Section 2, we review known facts for GCDT. In Section 3, our extended model is explained in detail. Both sections are almost separated by the two different subsections, Non-critical SFT Approach and Matrix Model Approach. At the end of Section 3, as a consistency check, we also consider the inclusive process, which turns out to reproduce our differential equation for disk amplitude. In Section 4, we discuss our model from the two different field theories, the non-critical SFT and the matrix model. 
      
\section{Generalized CDT}
\subsection{Non-critical SFT Approach}
We shall review the non-critical SFT of the original GCDT formulated in $\cite{a3}$. This model really reproduces the disk amplitude derived in the continuum limit of the strictly causal CDT in the case that the string coupling constant is zero. In this model, closed strings with length $L$ are created and annihilated from the vacuum, $|0\rangle$ ($\langle 0|$) by the operators, $\psi^{\dagger}(L)$ and $\psi(L)$, respectively:
\begin{equation}
\langle 0|\psi^{\dagger}(L)\ = \psi(L)|0\rangle = 0. \label{0}
\end{equation}
These creation and annihilation operators obey the following commutation relations:
\begin{equation}
[\psi(L), \psi^{\dagger}(L^{\prime})]=\delta (L-L^{\prime}), \label{1} 
\end{equation}
and the others are zero. The world-sheet which closed strings sweep out can be seen as the whole space-time itself.
Corresponding Hamiltonian can be written as:
\begin{align}
H_{0} &=\int^{\infty}_{0}dL\psi^{\dagger}(L) \mathcal{H}_{0}(L,\Lambda)\psi(L) +G_{s}\int^{\infty}_{0}dL_{1}\int^{\infty}_{0}dL_{2} \psi^{\dagger}(L_{1})\psi^{\dagger}(L_{2}) \psi(L_{1}+L_{2})(L_{1}+L_{2}) \notag \\
& \ \ \ +\alpha G_{s}\int^{\infty}_{0}dL_{1}\int^{\infty}_{0}dL_{2} \psi^{\dagger}(L_{1}+L_{2})\psi(L_{2}) \psi (L_{1})L_{2}L_{1} - \int^{\infty}_{0}dL\delta (L)\psi(L) \label{2}  ,
\end{align}
where
\begin{equation}
\mathcal{H}_{0}(L,\Lambda)=-L\partial^{2}_{L} + \Lambda L. \label{3} 
\end{equation}
 \begin{figure}[h]
 \begin{center}
   \includegraphics[width=100mm,clip]{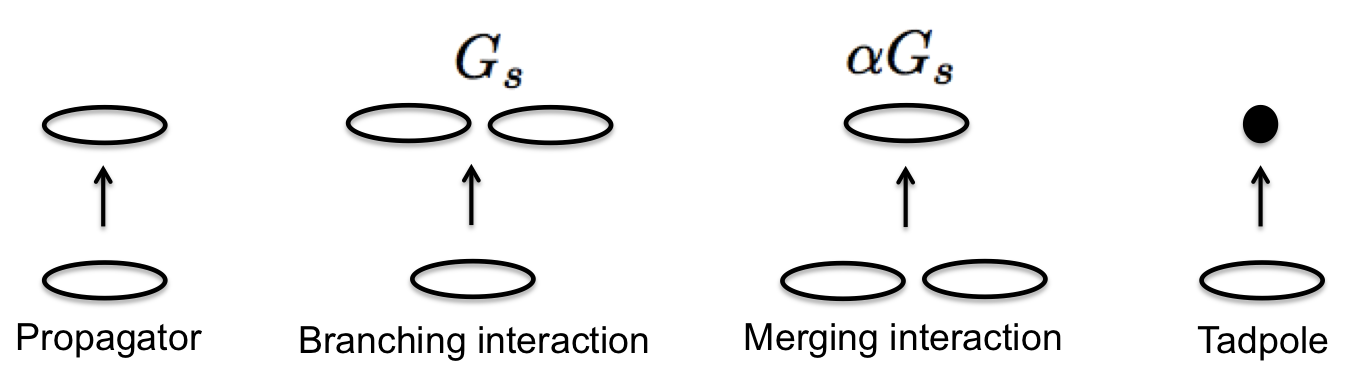}
  \caption{Terms in the Hamiltonian}\label{sft}
   \end{center}
\end{figure}
$G_{s}$ and $\Lambda$ are the string coupling constant and the cosmological constant, respectively. The parameter $\alpha$ in (\ref{2}) was introduced to count the numbers of genus in amplitudes. In the following discussion we shall take $\alpha =0$, which suppresses the creation of handles. The Hamiltonian above has been determined under the following scaling dimensions:
\begin{equation}
[S]=a, \ \ \ [\psi^{\dagger}(L)]=a^{0}, \ \ \ [\psi (L)]=a^{-1}, \ \ \ [G_{s}]=a^{-3}, \label{3.1}
\end{equation}
where $a$ is the scaling dimension of space, or alternatively speaking the lattice spacing, and $[S]$ is the scaling dimension of time.
A crucial difference between the Hamiltonian of the non-critical SFT constructed by Ishibashi and Kawai \cite{a1} and that of GCDT is the existence of the propagator term, $\int dL\psi^{\dagger}(L) \mathcal{H}_{0}\psi(L)$. In GCDT the propagator term actually exists but IK's theory does not. This difference comes from the fact that both theories have quite different definitions of ``time''.  

The authors in \cite{a3} derived Schwiner-Dyson's equation (SDE) for the Laplace-transformed disk amplitude, $\tilde{W}_{\Lambda}(Z)= \int^{\infty}_{0}dL e^{-LZ}  \langle0|e^{-SH_{0}}\psi^{\dagger}(L)|0\rangle |_{S\rightarrow \infty}$, in GCDT as \footnote{The authors derived the more general result with arbitrary $\alpha$, but here we restricted our situation to that with $\alpha =0$.}:
\begin{equation}
\partial_{Z} \bigl[ (\Lambda - Z^{2})\tilde{W}_{\Lambda}(Z) + G_{s} \tilde{W}_{\Lambda}^{2}(Z) \bigl] + 1=0. \label{3.1}
\end{equation} 
The solution of the above SDE was also derived by a perturbative expansion w.r.t. the string coupling constant in \cite{a3}:
\begin{equation}
\tilde{W}_{\Lambda}(Z) = \frac{1}{Z+\sqrt{\Lambda}} - G_{s}\frac{Z+3\sqrt{\Lambda}}{4\Lambda (Z+\sqrt{\Lambda})^{3}} + \mathcal{O}(G^{2}_{s}). \label{2.1.1}
\end{equation} 
The first term in the solution above is equivalent to the strictly causal solution \cite{a13}. In this formalism, the contributions from baby universes are weighted by the string coupling constant $G_{s}$.

\subsection{Matrix Model Approach}
The hermitian $N \times N$ matrix model reproducing the SDE of  GCDT was introduced. We start with the following matrix integral \cite{a9}:
\begin{equation}
\int d\phi \ e^{-\frac{N}{g_{s}} V(\phi) }, \label{3.3}
\end{equation}
where
\begin{equation}
V(\phi)=-g\phi + \frac{1}{2}\phi^{2} -\frac{1}{3}g\phi^{3}, \label{3.3.1}
\end{equation}
and $\phi$, $g$ and $g_{s}$ are a $N \times N$ hermitian matrix, the 'tHooft coupling constant and the string coupling constant, respectively. 
Then, by introducing the infinitesimal lattice spacing $a$, we can expand the coupling constants and the matrix w.r.t. $a$: 
\begin{equation}
g_{s}=\frac{1}{2}a^{3}G_{s}, \ \ \ \phi = \hat{I} -a \Phi +\mathcal{O}(a^{2}), \ \ \  g=\frac{1}{2}\biggl(1-\frac{1}{2}a^{2}\Lambda +\mathcal{O}(a^{4})\biggl), \label{3.4.1}
\end{equation}
where $\hat{I}$ is the unit $N \times N$ matrix, and $G_{s}$, $\Phi$ and $\Lambda$ are the corresponding renormalized values.
Substituting the fine-tuned values above into the potential $\frac{N}{g_{s}}V(\phi)$, we find
\begin{equation}
\frac{N}{g_{s}}\text{tr} V(\phi) = \frac{N}{G_{s}} \text{tr}\biggl( \frac{1}{3}\Phi^{3} - \Lambda \Phi \biggl) + (\text{terms independent of $\Phi$}) + \mathcal{O}(a). \label{3.4.2}
\end{equation}
Here we define the partition function in the continuum limit as:
\begin{equation}
Z \equiv \int d\Phi \exp \biggl[ {-\frac{N}{G_{s}}} \text{tr}\biggl( \frac{1}{3} \Phi^{3} - \Lambda \Phi \biggl) \biggl]. \label{3.4.4}
\end{equation}
In the large-$N$ limit, the saddle-point equation becomes\footnote{In \cite{a9}, the authors derived the general saddle-point equation beyond the large-$N$ limit. The general saddle-point equation really coincides with the SDE with arbitrary $\alpha$ by the treatment, $\alpha = 1/N^{2}$.}
\begin{equation}
\partial_{Z} \bigl[ (\Lambda - Z^{2})\tilde{W}_{\Lambda}(Z) + G_{s} \tilde{W}_{\Lambda}(Z)^{2} \bigl] +1=0, \label{3.5}
\end{equation}
where $\tilde{W}_{\Lambda}(Z)$ is the resolvent for the matrix $\Phi$. We notice that the saddle-point equation coincides with the SDE of GCDT.

\section{Generalized CDT with Extended Interactions}
\subsection{Non-critical SFT Approach}
Applying the method in \cite{a2}, we shall construct the non-critical SFT Hamiltonian of GCDT with extended interactions. 

The propagator term in $(\ref{2})$, $\int dL\psi^{\dagger}(L) \mathcal{H}_{0}(L,\Lambda)\psi(L)$, induces the strictly causal geometry. To make this propagator survive, we should impose the scaling dimension of space and time as:
\begin{equation}
[L]=a, \ \ \  [S]=a, \label{3.1}
\end{equation} 
where $a$ is the lattice spacing for space. From now, we shall extend the non-critical SFT based on GCDT without changing the scalings above. Since we think that the causality is the identity of CDT, this sort of extension is meaningful to get some deep understanding of what CDT is.    

First, we consider the strings with different charges, ($+$)-type and ($-$)-type. The creation and annihilation operators for ($+$)-type string, $\Psi^{\dagger}_{+}(L)$ and $\Psi_{+}(L)$, and for ($-$)-type string, $\Psi^{\dagger}_{-}(L)$ and $\Psi_{-}(L)$, are defined as the following vacuum conditions, respectively:
\begin{equation}
\langle 0|\psi_{+}^{\dagger}(L)\ = \psi_{+}(L)|0\rangle =\langle 0|\psi_{-}^{\dagger}(L)\ = \psi_{-}(L)|0\rangle = 0. \label{11}
\end{equation}
We assume these operators obey the following commutation relations: 
\begin{equation}
[\psi_{+}(L), \psi_{+}^{\dagger}(L^{\prime})]=[\psi_{-}(L), \psi_{-}^{\dagger}(L^{\prime})]=\delta (L-L^{\prime}),\label{12} 
\end{equation}
and the others are zero. Additionally, we assume the same scaling dimensions with those of GCDT:
\begin{equation}
[\psi^{\dagger}_{\pm}(L)]=a^{0}, \ \ \ [\psi_{\pm}(L)]=a^{-1}, \ \ \ [G_{s}]=a^{-3},
\end{equation}
where $G_{s}$ is the string coupling constant as before. 
Under the conditions above, we can extend the Hamiltonian for GCDT applying the interaction for spin clusters introduced by Ishibashi and Kawai \cite{a2}. Here we call such an interaction the \textit{IK-type interaction}. It is based on the so-called \textit{peeling procedure} in a discrete random surface. For example, considering a randomly triangulated surface coupled with Ising spins with one boundary and furthermore assuming that the boundary triangles have homogeneous spins (all spins are up-type or down-type), one peels triangles along with the boundary as if one peels an apple. If one continues to peel off triangles over the boundary triangles and one encounters the triangle having a different spin, then one surrounds the triangles having different spins by the triangles having same spins with the boundary triangles. In short, the randomly triangulated surface is separated by domain walls. In this case, the SDE for their approach coincides with the loop equation for the chain-type two-matrix model describing the random geometry coupled with Ising spins. We emphasize here that the above closed strings are not seen as the spin boundary as in the case of IK but the equal-time hypersurfaces with different charges.  
If we apply the IK-type interaction, we can write down the extended Hamiltonian for GCDT:
\begin{align}
H_{m} &=\int^{\infty}_{0}dL\psi^{\dagger}_{+}(L) \mathcal{H}_{0}(L,\Lambda)\psi_{+}(L) +G_{s}\int^{\infty}_{0}dL_{1}\int^{\infty}_{0}dL_{2} \psi^{\dagger}_{+}(L_{1})\psi^{\dagger}_{+}(L_{2}) \psi_{+}(L_{1}+L_{2})(L_{1}+L_{2}) \notag \\
&\ \ \  +b G_{s}\int^{\infty}_{0}dL_{1}\int^{\infty}_{0}dL_{2} \psi^{\dagger}_{+}(L_{1}+L_{2}) \psi^{\dagger}_{-}(L_{2}) \psi_{+}(L_{1})L_{1} \notag \\
&\ \ \ +\alpha G_{s}\int^{\infty}_{0}dL_{1}\int^{\infty}_{0}dL_{2} \psi^{\dagger}_{+}(L_{1}+L_{2}) \psi_{+}(L_{2}) \psi_{+}(L_{1})L_{2}L_{1} \notag \\
&\ \ \  - \int^{\infty}_{0}dL\delta (L)\psi_{+}(L) + \biggl[\psi_{+}( \ \psi_{+}^{\dagger}) \leftrightarrow  \psi_{-} ( \ \psi^{\dagger}_{-}) \biggl], \label{4}
\end{align}
\begin{figure}[h]
 \begin{center}
   \includegraphics[width=130mm,clip]{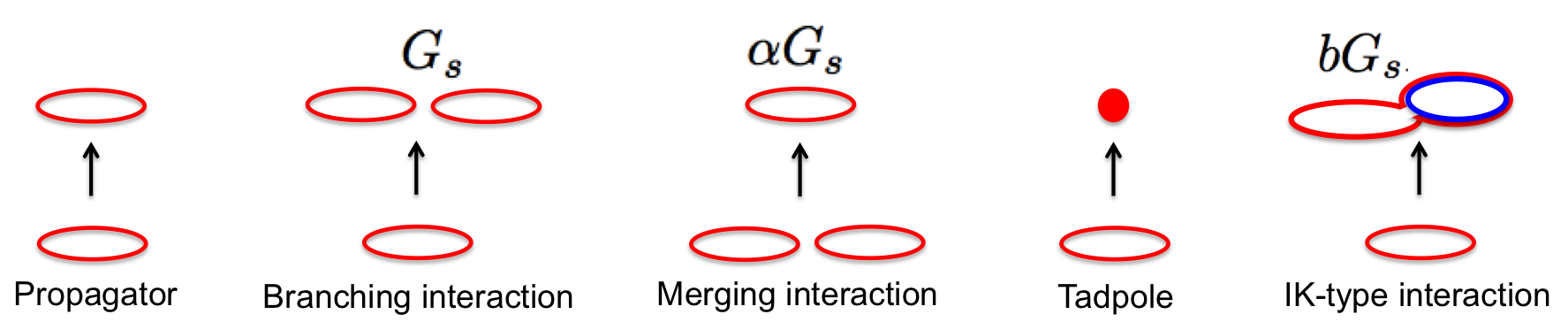}
  \caption{Terms in the extended Hamiltonian: The red string stands for the one having the ($+$)-type charge, and the blue for the ($-$)-type charge. Of course, terms whose charges are flipped exist in the Hamiltonian, but here we do not include the graphical expressions of those terms. }\label{exsft}
   \end{center}
 \end{figure}
where $\alpha$ and $b$ are dimension-less constants\footnote{In fact, it is possible to include the interactions, $\int dL\psi^{\dagger}_{-}(L) \mathcal{H}_{0}(L,\Lambda)\psi_{+}(L)$ and its spin-flipped term. However, because of the $\mathbb{Z}_{2}$-symmetry as to the spin reflection, such terms merely cause a constant shift of the string coupling constant, so that we have not included these terms in the Hamiltonian.}.     
In the Hamiltonian above, for simplicity, we will restrict the topology of geometries to that of a disk, which can be realized by the following Hamiltonian:
\begin{equation}
H_{m}^{D}\equiv \lim_{\alpha \rightarrow 0}H_{m}. \label{5}
\end{equation}

Next we will derive the SDE in our extended model. The SDE corresponds to Wheeler-DeWitt's equation for the wave function of the universe.
To begin, we define a partition function and disk amplitudes:
\begin{equation}
Z \equiv \lim_{S\rightarrow \infty} \langle0|e^{-SH^{D}_{m}}|0\rangle \equiv 1, \label{10}
\end{equation} 
and
\begin{equation}
W_{\pm}(L) \equiv \lim_{S\rightarrow \infty} \langle0|e^{-SH^{D}_{m}}\psi^{\dagger}_{\pm}(L)|0\rangle. \label{6}
\end{equation} 
The SDE for $W_{\pm}(L)$ is
\begin{equation}
\lim_{S\rightarrow \infty}\frac{\partial}{\partial S}\langle0|e^{-SH^{D}_{m}}\psi^{\dagger}_{\pm}(L)|0\rangle=0. \label{8}
\end{equation}
Using the equation, $H^{D}_{m}|0\rangle=0$, and the commutation relations (\ref{12}), we can rewrite the SDE as:
\begin{align}
0 &= -L\partial^{2}_{L}W_{\pm}(L)  + \Lambda LW_{\pm}(L) - \delta (L) +G_{s} L \int^{\infty}_{0}dL_{1}\lim_{S\rightarrow \infty} \langle0|e^{-SH^{D}_{m}}\psi^{\dagger}_{\pm}(L_{1})\psi^{\dagger}_{\pm}(L-L_{1})|0\rangle \notag \\
& \ \ \ +b G_{s}L \int^{\infty}_{0}dL_{1}\lim_{S\rightarrow \infty} \langle0|e^{-SH^{D}_{m}}\psi^{\dagger}_{\pm}(L+L_{1})\psi^{\dagger}_{\mp}(L+L_{1})|0\rangle . \label{13}
\end{align}
Here we introduce the factorization theorem:
\begin{align}
\lim_{S\rightarrow \infty} \langle0|e^{-SH^{D}_{m}}\psi^{\dagger}_{\pm}(L_{1})\psi^{\dagger}_{\pm}(L_{2})|0\rangle = \lim_{S\rightarrow \infty} \langle0|e^{-SH^{D}_{m}}\psi^{\dagger}_{\pm}(L_{1})|0\rangle \lim_{S\rightarrow \infty} \langle0|e^{-SH^{D}_{m}}\psi^{\dagger}_{\pm}(L_{2})|0\rangle. \label{14}
\end{align}
Applying the above factorization theorem, the SDE $(\ref{13})$ becomes
\begin{align}
0 &= -L\partial^{2}_{L} W_{\pm}(L) + \Lambda LW_{\pm}(L) - \delta (L) +G_{s}L \int^{\infty}_{0}dL_{1}W_{\pm}(L_{1})W_{\pm}(L-L_{1}) \notag \\
& \ \ \ + b G_{s}L \int^{\infty}_{0}dL_{1}W_{\pm}(L+L_{1})W_{\mp}(L_{1}) . \label{15}
\end{align}
In fact, our system has $\mathbb{Z}_{2}$-symmetry w.r.t. a spin-reflection, so that we will focus on a $\mathbb{Z}_{2}$-symmetric solution of the SDE: 
\begin{equation}
W_{+}(L)=W_{-}(L)\equiv W_{\Lambda}(L). \label{16}
\end{equation}

Next, we implement the Laplace transformation, $\mathcal{L}[W_{\Lambda}(L)]\equiv \int^{\infty}_{0}dL e^{-LZ}W_{\Lambda}(L)\equiv \tilde{W}_{\Lambda}(Z)$. Applying the expression, $W_{\Lambda}(L)$, and Laplace transforming (\ref{15}) yields
\begin{equation}
0 = \partial_{Z}\biggl[ (Z^{2}-\Lambda) \tilde{W}_{\Lambda}(Z) -G_{s}\tilde{W}_{\Lambda}(Z)^{2} \biggl] -1 + b G_{s}\mathcal{L}\biggl[L\int dL_{1}W_{\Lambda}(L+L_{1})W_{\Lambda}(L_{1})\biggl] . \label{18}
\end{equation}
We notice that the last term includes a divergent part as $Z\rightarrow \infty$. 
To regularize this divergence, it is good to symmetrize it w.r.t. the reflection, $Z\leftrightarrow -Z$ \cite{a2} \cite{ a4}:
\begin{equation}
\int^{\infty}_{0}dL\int^{\infty}_{0} dL_{1}e^{-Z(L+L_{1})}W_{\Lambda}(L+L_{1})e^{+ZL_{1}}W_{\Lambda}(L_{1}) + (Z\leftrightarrow -Z) = \tilde{W}_{\Lambda}(Z)\tilde{W}_{\Lambda}(-Z). \label{20}
\end{equation}
Subtracting the SDE with the reflection ($Z\rightarrow -Z$) from the original SDE (\ref{15}), we get the finite SDE:
\begin{equation}
0 = \partial_{Z}\biggl[ (Z^{2}-\Lambda)\left( \tilde{W}_{\Lambda}(Z)+\tilde{W}_{\Lambda}(-Z)\right) -G_{s}\left(\tilde{W}_{\Lambda}(Z)^{2} + \tilde{W}_{\Lambda}(-Z)^{2} + b \tilde{W}_{\Lambda}(Z)\tilde{W}_{\Lambda}(-Z)\right)  \biggl]  . \label{21}
\end{equation}  
Integration of the SDE above over $Z$ yields
\begin{equation}
c =  (Z^{2}-\Lambda)\left( \tilde{W}_{\Lambda}(Z)+\tilde{W}_{\Lambda}(-Z)\right) -G_{s}\left(\tilde{W}_{\Lambda}(Z)^{2} + \tilde{W}_{\Lambda}(-Z)^{2} + b \tilde{W}_{\Lambda}(Z)\tilde{W}_{\Lambda}(-Z)\right), \label{22}
\end{equation} 
where $c$ is a constant.

We will derive a perturbative solution for the SDE above around the weak coupling region, $G_{s}<1$, by expanding the loop amplitude $\tilde{W}_{\Lambda}(Z)$ and $c$ like:
\begin{equation}
\tilde{W}_{\Lambda}(Z)=\sum^{\infty}_{n=0}G^{n}_{s}W_{n}(Z), \ \ \ c=\sum^{\infty}_{n=0}G^{n}_{s}c_{n}. \label{22.1}
\end{equation}
As for $W_{0}(Z)$, we find
\begin{equation}
W_{0}(Z)  = \frac{1}{Z+\sqrt{\Lambda}}, \label{22.2}
\end{equation}
where we have chosen an overall constant for $W_{0}(Z)$ to coincide with that of pure CDT \cite{a13}. As for $W_{1}(Z)$ and $W_{1}(-Z)$, we find
\begin{equation}
W_{1}(Z)+W_{1}(-Z)= \frac{1}{(Z+\sqrt{\Lambda})^{3}(Z-\sqrt{\Lambda})^{3}}\biggl[ c_{1}Z^{4}+(2-b-2\Lambda c_{1})Z^{2} +c_{1} \Lambda^{2}+2\Lambda +b \Lambda \biggl]. \label{22.3}
\end{equation}
Assuming that the disk amplitude behaves as $1/Z$ in the large $Z$-region, we can determine that $c_{1} = -(b+1)/2\Lambda$. Furthermore, we can extract $W_{1}(Z)$ by considering that $W_{1}(Z)$ is analytic in the region, $\text{Re}[Z]>0$. Thus, the perturbative solution is
\begin{equation}
\tilde{W}_{\Lambda}(Z)=\frac{1}{Z+\sqrt{\Lambda}} - G_{s}\frac{1}{4\Lambda}\biggl[ \frac{Z+3\sqrt{\Lambda}}{(Z+\sqrt{\Lambda})^{3}} + \frac{b}{(Z+\sqrt{\Lambda})^{2}} \biggl] + \mathcal{O}(G^{2}_{s}). \label{22.4}
\end{equation}
The solution with $b =0$ is equivalent to that of the pure GCDT (\ref{2.1.1}).   
 
\subsection{Matrix Model Approach}
We start with the following matrix integral:
\begin{equation}
\int d\phi_{+}d\phi_{-}e^{-\frac{N}{g_{s}}V(\phi_{+}, \phi_{-})}, \label{23.1}
\end{equation}
where
\begin{equation}
V(\phi_{+},\phi_{-})=-g(\phi_{+} + \phi_{-}) +\frac{1}{2} (\phi_{+}^{2} + \phi_{-}^{2}) -\frac{g}{3}(\phi_{+}^{3} + \phi_{-}^{3}) +x\phi_{+}\phi_{-}. \label{23.2}
\end{equation}
In the integral above, $\phi_{\pm}$, $g$, $g_{s}$ and $x$ are $N\times N$ hermitian matrices, the 'tHooft coupling constant, the string coupling constant and the coupling constant characterizing the interaction, respectively.
Then, we expand the fields and coupling constants w.r.t. the lattice spacing $a$ as follows:
\begin{equation}
\phi_{+}=\hat{I} -a(A+B) + \mathcal{O}(a^{2}), \ \ \ \phi_{-}=\hat{I} -a(A-B) + \mathcal{O}(a^{2}), \label{23.3}
\end{equation}  
and
\begin{equation}
g_{s}=a^{3}G_{s}, \ \ \ g=\frac{1}{2}\biggl( 1-\frac{1}{2}a^{2}(\Lambda - 2X) +\mathcal{O}(a^{4})  \biggl), \ \ \ x=Xa^{2}, \label{23.4} 
\end{equation}
where $A$ and $B$ are $N \times N$ hermitian matrices, and $\hat{I}$ is the unit matrix, and $G_{s}$, $\Phi$, $\Lambda$ and $X$ are the corresponding renormalized values. 
Thus, our model can be seen as the one that the cut-length shrinks to zero ($g_{s}\rightarrow 0$), and the strength of the interaction falls off ($x \rightarrow 0$). 
The causality induces the scaling, $g_{s}\rightarrow 0$, and in addition, by taking the limit, $x \rightarrow 0$,  we can get our model as the weakly interacting model. 
Substituting the fine-tuned values, we can write down the partition function of the matrix model in the continuum limit:
\begin{equation}
Z=\int dAdB \exp \biggl[-\frac{N}{G_{s}}\text{tr}\biggl( \frac{1}{3}A^{3}+AB^{2}-\Lambda A \biggl) \biggl]. \label{23}
\end{equation} 
An interesting thing is that in the matrix model having this type of potential, the Gaussian integral over $B$ can be performed by introducing the eigenvalues $\lambda_{i}$'s for the matrix $A$ \cite{a4.1}: 
\begin{equation}
Z \propto \int\prod_{i}d\lambda_{i} \Delta^{2}(\lambda)\prod_{i,j}(\lambda_{i}+\lambda_{j})^{-1/2}e^{-\frac{N}{G_{s}}V}, \label{27}
\end{equation}
where
\begin{equation}
V=\sum_{i=1}^{N}V(\lambda_{i})=\sum_{i=1}^{N}\biggl( \frac{1}{3}\lambda^{3}_{i} - \Lambda \lambda_{i} \biggl), \label{27.1}
\end{equation}
and $\Delta (\lambda)$ denotes the Vandermonde determinant, $\Delta (\lambda)=\prod_{i<j}(\lambda_{j}-\lambda_{i})$. 
In the large-$N$ limit, the saddle point equation becomes 
\begin{equation}
\frac{2}{N}\sum_{j\neq i}\frac{1}{\lambda_{i}-\lambda_{j}}=\frac{1}{N}\sum_{j}\frac{1}{\lambda_{i}+\lambda_{j}}+\frac{1}{G_{s}}V^{\prime}(\lambda_{i}), \label{29}
\end{equation}  
where $V^{\prime}(\lambda_{i})=\lambda^{2}_{i}-\Lambda$.
Here we define the resolvent for $A$ as $\tilde{W}_{\Lambda}(Z) \equiv \frac{1}{N}\text{tr}(Z-A)^{-1}$, and the distribution of eigenvalues as $\rho(\lambda)\equiv \frac{1}{N}\sum_{i}\delta (\lambda - \lambda_{i})$.
Multiplying (\ref{29}) by $1/(Z-\lambda_{i})$ and summing over $i$, we obtain the loop equation in the large-$N$ limit:
\begin{equation}
V^{\prime}(Z)\tilde{W}_{\Lambda}(Z)+V^{\prime}(-Z)\tilde{W}_{\Lambda}(-Z)-G_{s} (\tilde{W}_{\Lambda}(Z)^{2}+\tilde{W}_{\Lambda}(Z)\tilde{W}_{\Lambda}(-Z)+\tilde{W}_{\Lambda}(-Z)^{2}) +G_{s}r_{1}(Z)=0, \label{33}
\end{equation} 
where
\begin{align}
G_{s}r_{1}(Z)&=\int d\lambda \rho (\lambda)\biggl[ \frac{V^{\prime}(\lambda)-V^{\prime}(Z)}{Z-\lambda} - \frac{V^{\prime}(\lambda)-V^{\prime}(-Z)}{Z+\lambda} \biggl] \notag \\
&= -2\int d\lambda \rho (\lambda) \lambda. \label{34}
\end{align}
In the calculation above, we used the two identities:
\begin{equation}
\frac{2}{N^{2}}\sum_{i\neq j}\frac{1}{Z-\lambda_{i}}\frac{1}{\lambda_{i}-\lambda_{j}}=\tilde{W}_{\Lambda}(Z)^{2}+\frac{1}{N}\tilde{W}_{\Lambda}^{\prime}(Z), \label{35}
\end{equation} 
and
\begin{equation}
\frac{1}{N^{2}}\sum_{i,j}\frac{1}{\lambda_{i}+\lambda_{j}}\biggl( \frac{1}{Z-\lambda_{i}}- \frac{1}{Z+\lambda_{i}} \biggl) =-\tilde{W}_{\Lambda}(Z)\tilde{W}_{\Lambda}(-Z). \label{36}
\end{equation} 
Putting explicit form of the potential into the loop equation (\ref{33}), we find
\begin{equation}
(Z^{2}-\Lambda)\left(\tilde{W}_{\Lambda}(Z)+\tilde{W}_{\Lambda}(-Z)\right) -G_{s} \left(\tilde{W}_{\Lambda}(Z)^{2}+\tilde{W}_{\Lambda}(Z)\tilde{W}_{\Lambda}(-Z)+\tilde{W}_{\Lambda}(-Z)^{2}\right) = 2\int d\lambda \rho (\lambda) \lambda. \label{37}
\end{equation} 
Remembering the SDE derived in the non-critical SFT approach (\ref{22}),
we can find a great similarity between the two. Namely, if we set $b =1$ in the SDE, then the two equations are exactly same. Thus, this matrix model in the continuum limit can reproduce our GCDT with extended interactions in $b=1$.  
   
We can extend the matrix model in the continuum limit above to the general $O(n)$ vector model \cite{a4.1} such that:
\begin{equation}
Z=\int dAdB_{1} \cdots dB_{n}e^{-\frac{N}{G_{s}}\text{tr}U(A,B_{1}, \cdots, B_{n})}, \label{38}
\end{equation} 
where
\begin{equation}
U(A,B_{1}, \cdots, B_{n})= A(B_{1}^{2} + \dots + B^{2}_{n}) + \frac{1}{3}A^{3} -\Lambda A, \label{39}
\end{equation} 
and $A$, $B_{1}$, \dots, $B_{n}$ are $N \times N$ hermitian matrices. 
One can find that the previous matrix model in the continuum limit is $O(1)$ vector model.
Again, we can integrate out all $B_{i}$'s, and a consequence is
\begin{equation}
Z\propto \int\prod^{N}_{i=1}d\lambda_{i}e^{-\frac{N}{G_{s}}V} \prod_{i,j}(\lambda_{i}+\lambda_{j})^{-n/2} \Delta^{2}(\lambda), \label{40}
\end{equation}
where $\lambda_{i}$s are eigenvalues of $A$, and $V=\sum_{i}V(\lambda_{i})=\sum_{i}(\frac{1}{3}\lambda^{3}_{i}-\Lambda \lambda_{i})$. A saddle-point equation becomes
\begin{equation}
\frac{2}{N}\sum_{j\neq i}\frac{1}{\lambda_{i}-\lambda_{j}}=\frac{n}{N}\sum_{j}\frac{1}{\lambda_{i}+\lambda_{j}}+\frac{1}{G_{s}}V^{\prime}(\lambda_{i}). \label{41}
\end{equation}  
In the similar manner as $O(1)$ vector model, we get the loop equation for the resolvent $\tilde{W}_{\Lambda}(Z)$:
\begin{equation}
(Z^{2}-\Lambda)(\tilde{W}_{\Lambda}(Z)+\tilde{W}_{\Lambda}(-Z)) -G_{s} (\tilde{W}_{\Lambda}(Z)^{2}+n\tilde{W}_{\Lambda}(Z)\tilde{W}_{\Lambda}(-Z)+\tilde{W}_{\Lambda}(-Z)^{2}) = 2\int d\lambda \rho (\lambda) \lambda. \label{42.1}
\end{equation}
Thus, the loop equation of this $O(n)$ vector model coincides with the SDE labeled by a free parameter $b$ (\ref{22}) only if we identify $n$ with $b$.  

\subsection{Inclusive Process}
In the above,  we derived the differential equation for disk amplitude in our extended model, and solved it by perturbative expansions. As a confirmation, we shall reproduce the same differential equation for disk amplitude using the so-called \textit{inclusive process} (\cite{a1}, \cite{a2} and \cite{a11}).     
In the inclusive process, putting caps (disk amplitudes) on one of two loops (universes) at branch points we can focus on the amplitude with one loop, which has its origin in the so-called \textit{transfer matrix formalism} \cite{a12}. If we focus on the case that initial and final strings have the same charges, then the inclusive SFT Hamiltonian can be written as follows:
\begin{align}
H_{IN} &=\int^{\infty}_{0}dL\psi^{\dagger}_{+}(L) \mathcal{H}_{0}(L,\Lambda)\psi_{+}(L) +2G_{s}\int^{\infty}_{0}dL_{1}\int^{\infty}_{0}dL_{2} W_{+}(L_{1})\psi^{\dagger}_{+}(L_{2}) \psi_{+}(L_{1}+L_{2})(L_{1}+L_{2}) \notag \\
&\ \ \ +b G_{s}\int^{\infty}_{0}dL_{1}\int^{\infty}_{0}dL_{2} \psi^{\dagger}_{+}(L_{1} +L_{2})W_{-}(L_{2})\psi_{+}(L_{1})L_{1} + \biggl[\psi_{+}( \ \psi_{+}^{\dagger}) \leftrightarrow  \psi_{-} ( \ \psi^{\dagger}_{-}) \biggl]. \label{43}
\end{align}  
Cylinder amplitudes are defined as:
\begin{equation}
G_{++}(L_{1},L_{2})\equiv \int^{\infty}_{0}dS G_{++}(L_{1},L_{2};S) \equiv \int^{\infty}_{0}\langle 0| \psi_{+}(L_{2}) e^{-SH_{IN}} \psi^{\dagger}_{+}(L_{1}) |0 \rangle , \label{43.1}
\end{equation}
\begin{equation}
G_{--}(L_{1},L_{2})\equiv \int^{\infty}_{0}dS G_{--}(L_{1},L_{2};S) \equiv \int^{\infty}_{0}\langle 0| \psi_{-}(L_{2}) e^{-SH_{IN}} \psi^{\dagger}_{-}(L_{1}) |0 \rangle. \label{44}
\end{equation}
By a differentiation of $G_{++}(L_{1},L_{2})$ w.r.t. time $S$, we find
\begin{equation}
\partial_{S}G_{++}(L_{1},L_{2};S) = - \langle 0| \psi_{+}(L_{2}) e^{-SH_{IN}}[H_{IN}, \psi^{\dagger}_{+}(L_{1})]|0 \rangle, \label{45}
\end{equation}
where we have used $H_{IN}|0\rangle =0$. With the calculation similar to (\ref{13}), the equation above can be rewritten as:
\begin{align}
\partial_{S}G_{++}(L_{1},L_{2};S) &= L_{1}(\partial^{2}_{L_{1}}-\Lambda )G_{++}(L_{1},L_{2};S) -2G_{s}L_{1} \int dL W_{+}(L_{1}-L)G_{++}(L,L_{2};S) \notag \\
                                                &\ \ \ -b G_{s}L_{1} \int dL W_{-}(L)G_{++}(L_{1}+L,L_{2};S) . \label{46.0}
\end{align}
Limiting the length $L_{2}$ to $0$ and integrating over $S$ in (\ref{46.0}), we have
\begin{equation}
0 = L_{1}(\partial^{2}_{L_{1}}-\Lambda )W_{+}(L_{1}) -2G_{s}L_{1} \int dL W_{+}(L_{1}-L)W_{+}(L) -b G_{s}L_{1} \int dL W_{-}(L)W_{+}(L_{1}+L), \label{46}
\end{equation}
where $W_{+}(L) \equiv \int^{\infty}_{0}dS G_{++}(L,0;S)$, and we have used the fact that $G_{++}(L_{1},0;\infty)=G_{++}(L_{1},0;0)=0$.
Our system has the $\mathbb{Z}_{2}$-symmetry as to the spin reflection, so that we focus on a $\mathbb{Z}_{2}$-invariant solution, $W_{\Lambda}(L)\equiv W_{\pm}(L)$, as in (\ref{16}).
Then, implementing the Laplace transformation of (\ref{46}) yields
\begin{equation}
0 = \partial_{X}\biggl[\left(-(X^{2}-\Lambda) + 2G_{s}\tilde{W}_{\Lambda}(X)\right) \tilde{W}_{\Lambda}(X) +b G_{s} \int dL_{1}e^{-XL_{1}} \int dL W_{\Lambda}(L)W_{\Lambda}(L_{1}+L) \biggl],\label{47}
\end{equation}
where $\tilde{W}_{\Lambda}(X)\equiv \int^{\infty}_{0}dLe^{-LX}W_{\Lambda}(L)$.
Again, the last term includes divergent part as $X\rightarrow \infty$. Thus, we need to remove the divergence by the symmetrization as in (\ref{20}):
\begin{align}
0&= \partial_{X}\biggl[(X^{2}-\Lambda )\left(\tilde{W}_{\Lambda}(X)+\tilde{W}_{\Lambda}(-X)\right) -2G_{s}\left( \tilde{W}_{\Lambda}(X)^{2} + \frac{b}{2}\tilde{W}_{\Lambda}(X)\tilde{W}_{\Lambda}(-X) + \tilde{W}_{\Lambda}(-X)^{2}\right) \biggl]. \label{49}
\end{align}
After the proper shifts of the string coupling constant $G_{s}$ and the free parameter $b$, the equation (\ref{49}) coincides with the SDE for the disk amplitude (\ref{22}) as expected.

\begin{figure}[h]
 \begin{center}
   \includegraphics[width=130mm,clip]{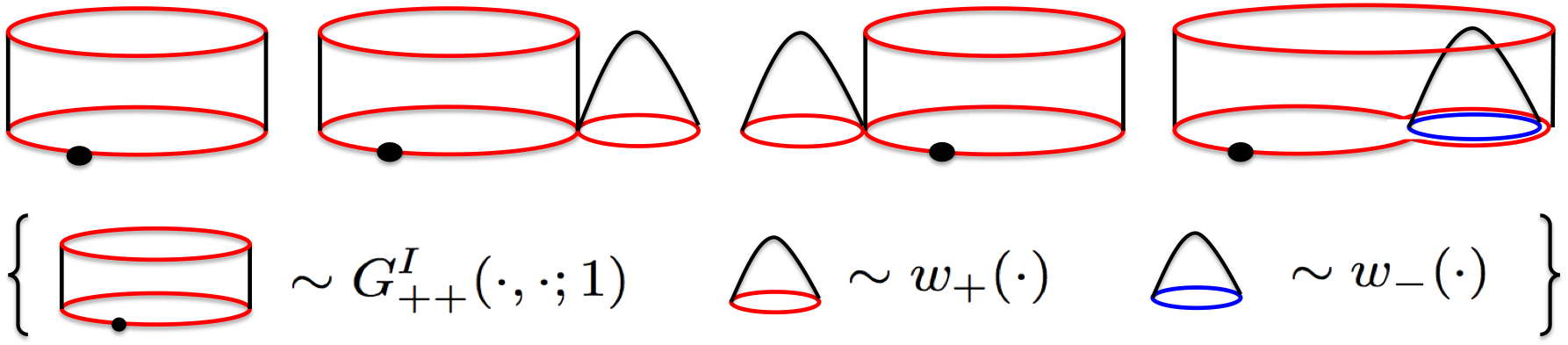}
  \caption{Terms in the transfer matrix, $G_{++}(l_{1},l_{2};1)$: dots in the arguments are replaced by some suitable variables on a case-by-case basis, and dots on loops in pictures are marked points on links.}\label{trm}
   \end{center}
\end{figure}

Next, we start with a discrete model, and then reconstruct  our model as its continuum limit. To carry it out, based on the transfer matrix formalism \cite{a12} we derive the differential equation for disk amplitude, which turns out to be equivalent to (\ref{22}) in the continuum limit.   
Here the transfer matrix is the one-time-step propagator having the length-$l_{1}$ initial loop with $(\pm)$-charge and the length-$l_{2}$ final loop with $(\pm)$-charge denoted by $G_{\pm \pm}(l_{1},l_{2};1)$. First, we derive the non-interacting propagator which is one of parts in the transfer matrix, $G_{\pm}^{I}(l_{1},l_{2};1)$. This can be easily calculated introducing the generating function of it, i.e. $\tilde{G}^{I}_{\pm \pm}(x_{\pm},y_{\pm};1) \equiv \sum_{l_{1},l_{2}}x^{l_{1}}_{\pm}y^{l_{2}}_{\pm}G^{I}_{\pm \pm}(l_{1},l_{2};1)$. Namely, we prepare four types of triangles weighted by $gx_{\pm}$ and $gy_{\pm}$ (Fig. \ref{weights}), and only from the combinatorics we can find the generating function of the one-time-step propagator \cite{a13}: 
\begin{figure}[h]
 \begin{center}
   \includegraphics[width=130mm,clip]{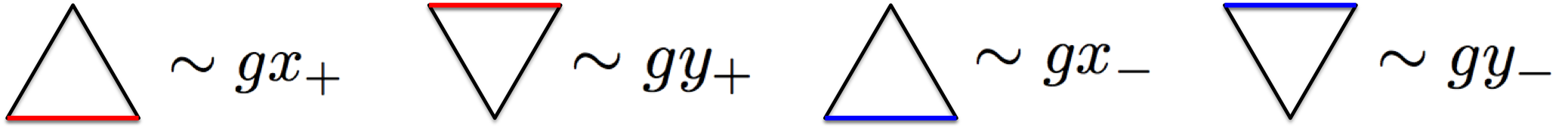}
  \caption{Four types of triangles weighted by $gx_{\pm}$ and $gy_{\pm}$: $gx_{\pm}$ and $gy_{\pm}$ are the weights for the triangles which are the elements of an initial loop with $(\pm)$-charge and of a final loop with $(\pm)$-charge in the generating function, respectively. }\label{weights}
   \end{center}
\end{figure}
\begin{equation}
\tilde{G}^{I}_{\pm \pm}(x_{\pm},y_{\pm};1)= \frac{g^{2}x_{\pm}y_{\pm}}{(1-gx_{\pm})(1-gx_{\pm}-gy_{\pm})}. \label{52.1}
\end{equation}
In the calculation above, we marked a point on one of initial links following \cite{a13}. For the later discussion, we give the specific form of weights:
\begin{equation}
g=\frac{1}{2}e^{-\frac{1}{2}a^{2} \Lambda}, \ \ \ x_{\pm}=e^{-aX_{\pm}}, \ \ \ y=e^{-aY_{\pm}}, \label{52.2} 
\end{equation} 
where $a$ is the lattice spacing, and $\Lambda$, $X_{\pm}$ and $Y_{\pm}$ are the renormalized coupling constants.  
Then, we introduce the transfer matrix combing the disk amplitude with the length-$l$ initial loop with ($\pm$)-charge $w_{\pm}(l)$ and the non-interacting propagator $G^{I}_{\pm \pm}(l_{1},l_{2};1)$ as follows:
\begin{align}
G_{\pm \pm}(l_1,l_2;1)&=G^{I}_{\pm}(l_1,l_2;1)+2g_s\sum_{l=1}^{l_1-1}l_1w_{\pm}(l_1-l)G^{I}_{\pm \pm}(l,l_2;1) +\hat{b}g_s\sum^{\infty}_{l=1}l_1w_{\mp}(l)G^{I}_{\pm \pm}(l+l_1,l_2;1), \label{52}
\end{align}   
where $g_{s}$ is the bare string coupling constant and $\hat{b}$ is a free parameter (Fig. \ref{trm}). 
A natural property of the propagator is the decomposition law:
\begin{equation}
G_{\pm \pm}(l_{1},l_{2};s) = \sum^{\infty}_{l=1}G_{\pm \pm}(l_{1},l;1)G_{\pm \pm}(l,l_{2};s-1). \label{54}
\end{equation}
After the Laplace transformation, the equation (\ref{54}) becomes
\begin{equation}
\tilde{G}_{\pm \pm}(x_{\pm},y_{\pm};s)=\oint \frac{dz_{\pm}}{2\pi iz_{\pm}}\tilde{G}_{\pm \pm}(x_{\pm},z_{\pm}^{-1};1)\tilde{G}_{\pm \pm}(z_{\pm},y_{\pm};s-1). \label{55}
\end{equation}
Substituting  (\ref{52.1}) into the equation (\ref{55}), one finds
\begin{align}
\tilde{G}_{\pm \pm}(x_{\pm},y_{\pm};s)&=\oint \frac{dz_{\pm}}{2\pi iz_{\pm}}\sum^{\infty}_{l_{1},l_{2}, l=1}x_{\pm}^{l_{1}}y_{\pm}^{l_{2}}z_{\pm}^{-l+l}G_{\pm \pm}(l_{1},l;1)G_{\pm \pm}(l,l_{2};s-1) \notag \\
&= \oint \frac{dz_{\pm}}{2\pi iz_{\pm}}\biggl[\tilde{G}^{I}_{\pm \pm}(x_{\pm},z_{\pm}^{-1};1)+2g_{s}x_{\pm}\partial_{x_{\pm}}\biggl(\tilde{w}_{\pm}(x_{\pm})\tilde{G}^{I}_{\pm \pm}(x_{\pm},z_{\pm}^{-1};1)\biggl) \notag \\
&\ \ \ + \sum^{\infty}_{l_{1}=1}\sum^{\infty}_{l^{\prime}=1}\hat{b} g_{s}x_{\pm}\partial_{x_{\pm}}\biggl(x_{\pm}^{l_{1}}w_{\mp}(l^{\prime}) \tilde{G}^{I}_{\pm \pm}(l^{\prime}+l_{1},z_{\pm}^{-1};1)\biggl)\biggl] \tilde{G}_{\pm \pm}\biggl(z_{\pm},y_{\pm};s-1\biggl), \label{55.1}
\end{align}
where $\tilde{w}_{\pm}(x_{\pm}) \equiv \sum_{l}x^{l}_{\pm}w_{\pm}(l)$, and $\tilde{G}^{I}_{\pm \pm}(l^{\prime}+l_{1},z^{-1}_{\pm};1) \equiv \sum_{l}z^{-l}_{\pm} G^{I}_{\pm \pm}(l^{\prime}+l_{1},l;1)$.
Here we introduce $l_{\text{cut}}$ to regularize the divergent summation over $l^{\prime}$ in (\ref{55.1}):
\begin{align} 
\tilde{G}_{\pm \pm}(x_{\pm},y_{\pm};s) &\rightarrow  \biggl[ 1+2g_{s}x_{\pm}(\partial_{x_{\pm}}\tilde{w}_{\pm}(x_{\pm}) + \tilde{w}_{\pm}(x_{\pm})\partial_{\pm} )\biggl] \frac{gx_{\pm}}{1-gx_{\pm}} \tilde{G}_{\pm \pm}      \biggl( \frac{g}{1-gx_{\pm}}, y_{\pm}; s-1  \biggl) \notag \\
&\ \ \ + \sum_{l_{1}=1}^{\infty} \sum^{l_{\text{cut}}}_{l^{\prime}=1}\hat{b}g_{s}x_{\pm}\partial_{x_{\pm}} \biggl[ x^{l_{1}}_{\pm}w_{\mp}(l^{\prime})\oint \frac{du_{\pm}}{2\pi i u_{\pm}}u^{-(l^{\prime}+l_{1})}_{\pm} \frac{gu_{\pm}}{1-gu_{\pm}}\tilde{G}_{\pm \pm}\biggl( \frac{g}{1-gu_{\pm}},y_{\pm}:s-1  \biggl)    \biggl],  \label{55.2}
\end{align} 
where $u_{\pm} \equiv e^{-aU_{\pm}}$.

In the following, we focus on $\mathbb{Z}_{2}$-symmetric solutions, i.e. $G_{\lambda}(x,y;s)\equiv G_{\pm \pm}(x_{\pm},y_{\pm};s)$ and $w_{\lambda}(x)\equiv w_{\pm}(x_{\pm})$.
Under the scalings, i.e., $S\equiv as$, $L_{1}\equiv al_{1}$ and $L_{2}\equiv al_{2}$, one finds the following renormalized functions:
\begin{equation}
\tilde{G}_{\Lambda}(X,Y;S)=\lim_{a\rightarrow 0}a\tilde{G}_{\lambda}(x,y;s), \ \ \tilde{W}_{\Lambda}(X)=\lim_{a\rightarrow 0}a\tilde{w}_{\lambda}(x), \ \ W_{\Lambda}(L)=\lim_{a \rightarrow \infty} w_{\lambda}(l). \label{61}
\end{equation}
From (\ref{52.2}), (\ref{55.1}), (\ref{55.2}) and (\ref{61}), one finds
\begin{align}
\partial_{S}\tilde{G}_{\Lambda}(X,Y;S) &= -\partial_{X} \biggl[ (X^{2}-\Lambda )\tilde{G}_{\Lambda}(X,Y;S) +2G_{s}\tilde{W}_{\Lambda}(X)\Tilde{G}_{\Lambda}(X,Y;S)   \notag \\
& \ \ \ + \lim_{L_{\text{cut} \rightarrow \infty}} \hat{b}G_{s}   \int^{\infty}_{0}dL_{1}  \int^{L_{\text{cut}}}_{0}dL^{\prime} e^{L^{\prime}X}W_{\Lambda}(L^{\prime}) e^{-(L^{\prime} + L_{1})X}\tilde{G}_{\Lambda}(L^{\prime}+L_{1},Y;S) \biggl], \label{59}
\end{align}
where $L_{\text{cut}} \equiv al_{\text{cut}}$, and we have used $\tilde{G}_{\Lambda}(L^{\prime}+L_{1},Y;S) \equiv \int^{i\infty}_{-i\infty}dU e^{(L^{\prime}+L_{1})U} \tilde{G}_{\Lambda}(U,Y;S) $. 
Implementing the inverse Laplace transformation w.r.t. $Y$ and integrating over $S$ in (\ref{59}), one finds
\begin{align}
0 &= -\partial_{X} \biggl[ (X^{2}-\Lambda )\tilde{G}_{\Lambda}(X,L) +2G_{s}\tilde{W}_{\Lambda}(X)\Tilde{G}_{\Lambda}(X,L)  \notag \\
&\ \ \ +  \hat{b}G_{s}   \int^{\infty}_{0}dL_{1}  \int^{\infty}_{0}dL^{\prime} e^{L^{\prime}X}W_{\Lambda}(L^{\prime}) e^{-(L^{\prime} + L_{1})X}G_{\Lambda}(L^{\prime}+L_{1},L) \biggl], \label{100}
\end{align}
where $\tilde{G}_{\Lambda}(X,L) \equiv \int^{\infty}_{0}dS \tilde{G}_{\Lambda}(X,L;S)$, and $G_{\Lambda}(L^{\prime}+L_{1},L) \equiv \int^{i \infty}_{-i \infty}dYe^{LY}\tilde{G}_{\Lambda}(L^{\prime}+L_{1},Y)$.
Limiting the length $L$ to 0 in (\ref{100}), one finds
\begin{align}
0 &= \partial_{X} \biggl[ (X^{2}-\Lambda )\tilde{W}_{\Lambda}(X) +2G_{s}\tilde{W}_{\Lambda}(X)^{2}  +  \hat{b}G_{s}   \int^{\infty}_{0}dL_{1}  \int^{\infty}_{0}dL^{\prime} e^{L^{\prime}X}W_{\Lambda}(L^{\prime}) e^{-(L^{\prime} + L_{1})X}W_{\Lambda}(L^{\prime}+L_{1}) \biggl], \label{101}
\end{align}
where $\tilde{W}_{\Lambda}(X)\equiv \tilde{G}_{\Lambda}(X,0)$ and $W_{\Lambda}(L^{\prime}+L_{1})\equiv G_{\Lambda}(L^{\prime}+L_{1},0)$. 
As for the last term in (\ref{101}), we use the same procedure as in (\ref{20}) and (\ref{49}):
\begin{equation}
 \int^{\infty}_{0}dL_{1}\int^{\infty}_{0}dL^{\prime} e^{L^{\prime}X}W_{\Lambda}(L^{\prime}) e^{-(L^{\prime}+L_{1})X} W_{\Lambda}(L^{\prime}+L_{1}) +(X \leftrightarrow -X)  = \tilde{W}_{\Lambda}(-X)\tilde{W}_{\Lambda}(X). \label{58.1}
\end{equation} 
Therefore, we obtain the finite differential equation for disk amplitude:
\begin{align}
0&= \partial_{X}\biggl[(X^{2}-\Lambda )\left(\tilde{W}_{\Lambda}(X)+\tilde{W}_{\Lambda}(-X)\right) +2G_{s}\left( \tilde{W}_{\Lambda}(X)^{2} + \frac{\hat{b}}{2}\tilde{W}_{\Lambda}(X)\tilde{W}_{\Lambda}(-X) + \tilde{W}_{\Lambda}(-X)^{2}\right) \biggl] \label{103}
\end{align}
Finally, after the proper shifts of the string coupling constant $G_{s}$ and the free parameter $\hat{b}$, one finds that (\ref{103}) is equivalent to (\ref{22}) as expected.  

 
\section{Discussions}
We have shown the equivalence between the two different field theories at the level of differential equations, the Schwinger-Dyson's equation in the non-critical SFT and the loop equation of the matrix model in the continuum limit. We hope that our model is a first step toward matter-coupled systems based on CDT. In the following, we will examine the model constructed in this paper from different point of view. 

To begin with, we will discuss our model in terms of the SFT approach. Although we have used the IK-type interaction to construct the extended SFT based on GCDT, we do not understand whether or not our model is on the critical point of the Ising model characterized by the Curie temperature. In the following, we will explain two complications around this problem. First, at the critical point of Ising spins the spin configuration must be random. In other words, the spins are supposed to fluctuate all length scales between the lattice spacing and the correlation length. Contrary to that, in our model the homogeneous spin (charge)  configurations survive as the propagators. Second, the definition of time induced by our Hamiltonian (\ref{4}) is different from the would-be GCDT coupled with Ising spins. Namely, we consider the closed strings in our model as not spin-cluster boundaries but spacial boundaries, so that we pursue the time flow of spatial boundaries. Thus, our time is nothing but the proper time. This proper time is crucially different from the time defined via the spin-cluster boundary \cite{a21} \cite{a22}. If we consider our time as the one defined via the spin-cluster boundary, which is equivalent to treating our model as the GCDT coupled with Ising spins, then the scaling dimension of time may be different from the lattice spacing $a$ according to \cite{a21}. This contradicts our first setup (\ref{3.1}).          
Anyhow, the free parameter $b$ might be the key to know what our model is. 

In addition, it is possible to extend our non-critical SFT to the multi-``colored'' system: 
\begin{align}
H^{(n)}_{m} &=\sum^{n}_{i=1}\int^{\infty}_{0}dL\psi^{\dagger}_{i}(L) \mathcal{H}_{0}(L,\Lambda)\psi_{i}(L) +G_{s} \sum^{n}_{i=1}  \int^{\infty}_{0}dL_{1}\int^{\infty}_{0}dL_{2} \psi^{\dagger}_{i}(L_{1})\psi^{\dagger}_{i}(L_{2}) \psi_{i}(L_{1}+L_{2})(L_{1}+L_{2}) \notag \\
&\ \ \ + G_{s}\sum^{n}_{i=1}\sum^{n}_{j \neq i}b_{ij}\int^{\infty}_{0}dL_{1}\int^{\infty}_{0}dL_{2} \psi^{\dagger}_{i}(L_{1}+L_{2}) \psi^{\dagger}_{j}(L_{2}) \psi_{i}(L_{1})L_{1} \notag \\
&\ \ \ +\alpha G_{s}\sum^{n}_{i=1}\int^{\infty}_{0}dL_{1}\int^{\infty}_{0}dL_{2} \psi^{\dagger}_{i}(L_{1}+L_{2}) \psi_{i}(L_{2}) \psi_{i}(L_{1})L_{2}L_{1} \notag \\
&\ \ \ -  \sum^{n}_{i=1}\int^{\infty}_{0}dL\delta (L)\psi_{i}(L). \label{}
\end{align}
We can derive the free parameter $b$ in our model from the multi-``colored'' system above under the treatment, $W_{1}(L)=\cdots W_{n}(L)\equiv W_{\Lambda}(L)$, $b_{ij}=0$ for $j=i$ and $b_{ij}=1$ for $j \neq i$.     

Next, we will closely look at our matrix model.
Considering the direct product of the two copies of the potential, each of which yields the pure GCDT, and introducing the linear combinations of the matrices as $\Phi_{+}=A+B$ and $\Phi_{-}=A-B$, we find 
\begin{equation}
\frac{1}{\tilde{G_{s}}}\biggl( \frac{1}{3}\Phi_{+}^{3} - \Lambda \Phi_{+} + \frac{1}{3}\Phi^{3}_{-} - \Lambda \Phi_{-} \biggl)=\frac{1}{G_{s}}\biggl( \frac{1}{3}A^{3} +AB^{2} -\Lambda A \biggl), 
\end{equation}
where $\tilde{G_{s}}=2G_{s}$. This is the potential of our $O(1)$ vector model in the continuum limit. Then, diagonalizing the matrix $A$ as $A=\text{diag}(\lambda_{1}, \cdots ,
\lambda_{N})$ and integrating out the matrix $B$, we get the effective theory for the eigenvalues of $A$ with the potential, 
\begin{equation}
\underbrace{ \biggl[ \frac{1}{G_{s}}\sum_{i}\biggl(\frac{1}{3}\lambda_{i}^{3}  -\Lambda \lambda_{i}\biggl) -\frac{1}{N}\log \Delta^{2} (\lambda)\biggl]}_{\text{terms appeared in the pure GCDT}} + \frac{1}{N}(\text{terms induced by the integration over $B$}). 
\end{equation}
The important point here is that our model is slightly different from the pure GCDT matrix model because integrating out the matrix $B$ an extra correction is added to terms appeared in the pure GCDT. From the matrix $A$'s point of view, the matrix $B$ can be seen like some external field. The strength of such an external field can be bigger by inserting the integrated-out matrices, which leads to the $O(n)$ vector model in the continuum limit.  

\section*{Acknowledgements}
The authors are very grateful to Jan Ambj$\o$rn. Thanks to his comments on the early stage of this work, we were able to understand our model deeply. One of the authors, YS also thanks Bergfinnur Durhuus, Shinji Hirano, Noboru Kawamoto, Charlotte Kristjansen, Sanefumi Moriyama, Ryuichi Nakayama and Tadakatsu Sakai for fruitful comments and discussions. YS thanks the warm hospitality of Niels Bohr Institute.   
Two of the authors, HF and YS, are supported by the Grant-in-Aid for Nagoya University Global COE Program, ``Quest for Fundamental Principles in the Universe: from Particles to the Solar System and the Cosmos.'' The work of HF is also supported by the Grant-in-Aid for Young Scientists (B) [\# 21740179] from the Japan Ministry of Education, Culture, Sports, Science and Technology. 

\pagebreak

\end{document}